\documentclass[amsmath,twocolumn,showpacs,letterpaper,floatfix]{revtex4}
\usepackage{amssymb}
\usepackage{graphicx}

\allowdisplaybreaks[1]

\begin{document}

\title{High-Precision Thermodynamic and Critical Properties\\
from Tensor Renormalization-Group Flows} \author{Michael
Hinczewski$^{1}$ and A. Nihat Berker$^{1-3}$}

\affiliation{$^1$Feza G\"ursey Research Institute, T\"UBITAK -
Bosphorus University, \c{C}engelk\"oy 34684, Istanbul, Turkey,}

\affiliation{$^2$Department of Physics, Ko\c{c} University, Sar\i
yer 34450, Istanbul, Turkey,}

\affiliation{$^3$Department of Physics, Massachusetts Institute of
Technology, Cambridge, Massachusetts 02139, U.S.A.}

\begin{abstract}
The recently developed tensor renormalization-group (TRG) method
provides a highly precise technique for deriving thermodynamic and
critical properties of lattice Hamiltonians.  The TRG is a local
coarse-graining transformation, with the elements of the tensor at
each lattice site playing the part of the interactions that undergo
the renormalization-group flows.  These tensor flows are directly
related to the phase diagram structure of the infinite system, with
each phase flowing to a distinct surface of fixed points.
Fixed-point analysis and summation along the flows give the critical
exponents, as well as thermodynamic functions along the entire
temperature range.  Thus, for the ferromagnetic triangular lattice
Ising model, the free energy is calculated to better than $10^{-5}$
along the entire temperature range.  Unlike previous position-space
renormalization-group methods, the truncation (of the tensor index
range $D$) in this general method converges under straightforward
and systematic improvements.  Our best results are easily obtained
with $D=24$, corresponding to 4624-dimensional renormalization-group
flows.

\end{abstract}

\maketitle

\section{Introduction}

The tensor renormalization-group (TRG) method, recently introduced
by Levin and Nave \cite{LevinNave} is versatile, accurate, and
conceptually interesting.  The versatility of TRG comes from being a
genuinely local renormalization-group transformation, that is a
mapping between local Hamiltonians on the original and
coarse-grained lattices, expressed in terms of tensors at sites of
each lattice.  Thus, although TRG has been demonstrated by obtaining
very accurate phase transition temperatures and magnetization curves
for frustrated and unfrustrated classical Ising models on the
triangular lattice, we believe that it can be extended to yield the
critical properties and entire thermodynamics of complicated
multicritical systems, including quenched random systems, by
renormalization-group flow analysis.

In the present work, we show that the renormalization-group flows of
the tensor elements can be analyzed in the same manner as the flows
of Hamiltonian interaction parameters, which was not done in the
previous work.  Each thermodynamic phase region corresponds to a
basin of attraction in the space of tensor elements, and critical
temperatures and exponents (not obtained before) are deduced from
characteristics of the boundaries between the different basins.
Practically, this means that the phase diagram, thermodynamic, and
critical properties of the infinite system can be derived directly
from the flows. Moreover, the flows themselves have an interesting
and unconventional nature, since a phase region does not flow to
single isolated sink, but rather to a continuous surface of fixed
points. Using the ferromagnetic triangular lattice Ising model as an
example, we show that the accuracy of the calculated free energy,
critical temperature, and thermal critical exponent can be easily
and systematically improved on by increasing $D$, the cutoff on the
index range of the tensors. In this ability to converge toward exact
values with larger cutoffs, TRG is more general and dependable than
position-space renormalization-group techniques
\cite{NvL,Migdal,Kadanoff} whose successes have been based on
system-specific heuristics \cite{Kadanoff1,Kadanoff2,Berker}. The
TRG approach combines the straightforward interpretative framework
of traditional renormalization group---analysis of flows in a
parameter space---with the accuracy of techniques depending on
finite-size scaling of large systems.

\section{Tensor renormalization-group transformation}

The TRG method \cite{LevinNave} can be applied to any classical
lattice Hamiltonian satisfying the following conditions: (1) it can
be expressed in terms of degrees of freedom on the bonds of the
lattice; (2) the Boltzmann weight of a configuration can be written
as the product of individual Boltzmann weights for each lattice
site, that only depend on the bond variables adjoining the site.
This is a broad category including many common statistical physics
systems like the Ising and Potts models, both of which can be mapped
onto this form through duality transformations, as well as all
vertex models.  For such Hamiltonians, the partition function can be
written as a {\it tensor network} \cite{MarkovShi,ShiDuanVidal}: We
start with a lattice of $N$ sites where each site has coordination
number $q$ and each bond is in one of $d$ possible states. The
Boltzmann weight of an individual site depends on the configuration
of the $q$ bonds meeting at the site and can be written as a tensor
$T_{i_1 i_2 \cdots i_q}$, where each index $i_\alpha$ runs from 1 to
$d$. For configurations of bond variables that are not allowed, the
corresponding element of $T$ is zero. The tensor is real-valued and
cyclically symmetric.  The transformation described below works also
for the more general case of complex-valued, cyclically symmetric
tensors, onto which our original tensor network system is mapped
after a single renormalization step. The partition function is a
product over the $N$ site tensors, with each index contracted
between two different tensors (since each bond is shared between two
sites),
\begin{equation}\label{eq:2}
Z = \sum_{i_1,\ldots,i_M=1}^d T_{i_1 i_2\cdots i_q} T_{i_1 i_r \cdots
i_s} T_{i_2 i_t\cdots i_u}\:\cdots\,.
\end{equation}

The position-space renormalization-group transformation of this
system allows us to express the partition function equivalently as a
tensor network over a coarse-grained lattice with $N^\prime = N/b^2$
vertices, in two spatial dimensions, with a length rescaling factor
$b>1$. It is accomplished in two steps, which we call rewiring and
decimation.  For simplicity, we shall focus in this work on applying
the transformation to the hexagonal lattice, though this two-step
procedure can easily be adapted to a variety of two-dimensional
lattices, including the square~\cite{LevinNave} and Kagome lattices.

\begin{figure}
\includegraphics[scale=1]{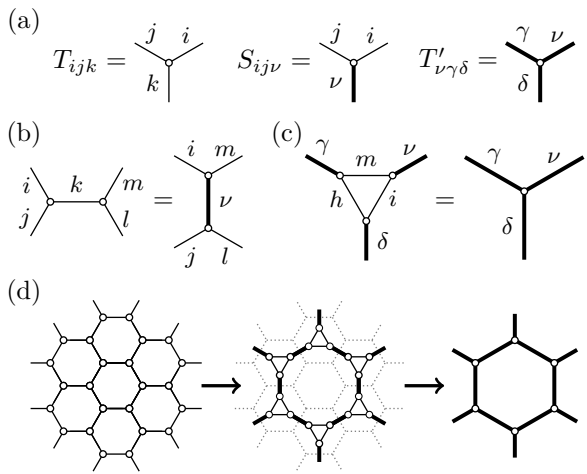}
\caption{(a) Graphical representation of the tensors.  (b) Rewiring.
(c) Decimation.  (d) Renormalization group transformation applied to
the entire hexagonal lattice, with the first arrow showing the
rewiring step, and the second arrow showing the
decimation.}\label{fig1}
\end{figure}

{\it Rewiring:} Graphically, let us represent the tensor $T_{ijk}$
at each point of the hexagonal lattice as a three-legged vertex as
shown on the left of Fig. \ref{fig1}(a), and a contraction of an
index between two tensors as a connection between two vertex legs.
We use the convention that the order of indices in the tensor
matches the counterclockwise ordering of the labels on the vertex
legs.  The rewiring step for a pair of neighboring tensors consists
of reconnecting the bonds in the manner of Fig. \ref{fig1}(b),
rewriting them as a contraction of two new tensors $S$,
\begin{equation}\label{eq:3}
\sum_{k=1}^d T_{ijk} T_{klm} = \sum_{\nu=1}^{d^2} S_{mi\nu}S_{jl\nu}\,.
\end{equation}
The first two indices in the tensor $S_{mi\nu}$ run up to $d$, but
the third index $\nu$ runs up to $d^2$.  We shall distinguish these
types of indices by a Greek letter label, and graphically depict
them as thick legs [Fig. \ref{fig1}(a) center].  To see that such a
rewiring is possible, let us introduce composite indices $\alpha
\equiv (m,i)$ and $\beta \equiv (j,l)$, and write the tensor
contraction on the left-hand side of Eq. \eqref{eq:3} as a $d^2
\times d^2$ matrix $M_{\alpha\beta} \equiv \sum_k T_{ijk}T_{klm}$.
Then Eq. \eqref{eq:3} becomes
\begin{equation}\label{eq:4}
M_{\alpha\beta} = \sum_\nu S_{\alpha\nu} S_{\beta\nu},
\end{equation}
or $M = S S^T$.  We can find $S$ using the fact that $M = M^T$, as
can be checked using the cyclical symmetry of the $T$ tensors.  Any
symmetric matrix $M$ admits a variant of singular value
decomposition known as Takagi factorization~\cite{Takagi}: $M = U
\Sigma U^T$, where $U$ is a $d^2 \times d^2$ unitary square matrix
and $\Sigma$ is a $d^2 \times d^2$ diagonal matrix containing the
singular values of $M$.  If $\Sigma_{\nu\nu}$ is the $\nu$th
singular value (assumed ordered from largest to smallest with
increasing $\nu$), then the elements of $S$ are given by
$S_{\alpha\nu} = \sqrt{\Sigma_{\nu\nu}} U_{\alpha\nu}$. The
resulting decomposition $M= S S^T$ is not unique, because we can
replace $S$ by $SO$, where $O$ is a complex orthogonal matrix ($O
O^T = I$).  However, we shall always use the $S$ given directly by
the the Takagi factorization, so if the $\nu$th singular value
$\Sigma_{\nu\nu}$ is nondegenerate, the $\nu$th column of $S$ is
uniquely determined up to a factor of $\pm 1$.  The rewiring
procedure described above is applied globally to the entire
hexagonal lattice as shown by the first arrow in Fig. \ref{fig1}(d),
grouping the $T$ tensors into pairs, and replacing each pair by $S$
tensors.

{\it Decimation:} The second step in the renormalization procedure
consists of tracing over the degrees of freedom in the triangular
clusters that are formed after the rewiring.  As illustrated in Fig.
\ref{fig1}(c), each such cluster can be replaced by a single
renormalized tensor $T^\prime$,
\begin{equation}\label{eq:5}
\sum_{m,i,h=1}^d S_{hm\gamma} S_{mi\nu} S_{ih\delta} =
T^\prime_{\gamma\delta\nu}\,.
\end{equation}
Introducing the notation $S^{(\gamma)}$ for the $d \times d$ matrix
with elements $S_{hm\gamma}$ at fixed $\gamma$, then the above
equation has the form
\begin{equation}\label{eq:6}
\text{Tr}[S^{(\gamma)} S^{(\nu)} S^{(\delta)}] = T^\prime_{\gamma\delta\nu}\,.
\end{equation}
It is clear from Eq. \eqref{eq:6} that the renormalized tensor will
be cyclically symmetric.  Since the $S$ tensor is constructed from
the unitary matrix $U$, the elements of $T^\prime$ will in general
be complex. The second arrow of Fig. \ref{fig1}(d) shows the
decimation applied to every triangular cluster, and the result is a
hexagonal lattice of $T^\prime$ tensors with the lattice spacing
larger by a factor of $b=\sqrt{3}$ compared to the original
hexagonal lattice.

The renormalization-group transformation described so far is an
exact mapping of one tensor network onto another, preserving the
partition function $Z$. However, the renormalized tensors have a
more complicated structure than the original ones, since the indices
of $T^\prime_{\gamma\delta\nu}$ run up to $d^2$.  If this procedure
is repeated, the index range grows exponentially with iteration
number, making numerical implementation impractical.  This is
analogous to the difficulty encountered when applying naive
position-space renormalization to Hamiltonians on lattices, where
the number of couplings in the Hamiltonian grows with each
iteration.  Some kind of approximate truncation is required to keep
the complexity of the renormalized model bounded.  In our case this
truncation can be done in a straightforward and systematic fashion
by setting an upper bound $D$ for the index range. Rather than use
the full $d^2 \times d^2$ matrix $S_{\alpha\nu}$ to calculate
$T^\prime$, we use only the first $\bar{d}$ columns, where $\bar{d}
\equiv \min(d^2,D)$.  Using this truncated $d^2 \times \bar{d}$
matrix, which we call $\bar{S}$, means that the rewiring step is
implemented only approximately, $M \approx \bar{S} \bar{S}^T$.
However, this is the best approximation possible, since the first
$\bar{d}$ columns of $S$ correspond to the $\bar{d}$ largest
singular values of $M$. The decimation step of Eq. \eqref{eq:5} is
still carried out exactly, but with $S$ replaced by $\bar{S}$.  The
resulting tensor $T^\prime_{\gamma\delta\nu}$ has indices that run
up to $\bar{d}$.

\section{Renormalization-Group Flows and Thermodynamic Behavior of the Triangular Lattice Ising Model}

In order to illustrate the nature of the renormalization-group flows
resulting from the transformation outlined above and the methods by
which thermodynamic information can be extracted from them, we turn
now to a specific example: the triangular lattice ferromagnetic
Ising model.  This model is mapped onto a tensor network through a
duality transformation, giving an equivalent partition function in
terms of bond variables on a hexagonal lattice.\cite{LevinNave}
Each bond in this dual lattice has two states, $\sigma_1 = 1$ or
$\sigma_2 = -1$, corresponding to a bond between parallel or
antiparallel Ising spins.  These states respectively contribute
energies $-J<0$ and $J$ to the total Hamiltonian $\mathcal{H}$.  The
tensor for each lattice site has the form
\begin{equation}\label{eq:7}
T_{ijk} = e^{\frac{\beta J}{2}(\sigma_i + \sigma_j + \sigma_k)}\frac{1}{2}(\sigma_i \sigma_j \sigma_k +1)\,,
\end{equation}
where $\beta = 1/k_B T$, and the factor multiplying the exponential
is a projection operator that is equal to $1$ for allowed
configurations of the bonds and $0$ otherwise.  For simplicity we
set $J/k_B=1$, effectively measuring temperatures in units of
$J/k_B$. If the triangular lattice has $N$ sites, the dual hexagonal
lattice has $2N$ sites, and the partition function is a contraction
of the $2N$ site tensors.

Iterating the renormalization-group transformation, we consider
renormalization-group flows in the space of tensor element
amplitudes $|T_{\alpha \beta \gamma}|$. For a cyclically symmetric
tensor $T_{ijk}$ with index range $d$, the maximum number of
distinct elements is $d(2+d^2)/3$. After the first few iterations,
the index range of the renormalized tensors reaches the cutoff $D$
and remains there, so that the flows in the subsequent steps are in
a space with dimension $D(2+D^2)/3$.  In this work we look at
cutoffs in the range $D = 4$ to $24$, so the most complex flow space
that we investigate is 4624-dimensional.  A large number of the
$D(2+D^2)/3$ tensor elements are zero, corresponding to disallowed
bond configurations, typically making up between 50-55\% of the
total in the later steps. The locations of the zero elements usually
stay the same for a few steps at a time, but can shift abruptly
during the flow.  To minimize roundoff errors which can affect the
results at later iterations, the singular value decomposition and
matrix manipulations involved in TRG are all implemented at
quadruple precision, using a customized version of the LAPACK
library~\cite{LAPACK}.

Due to the contraction during the decimation step, the tensor
elements tend to increase in magnitude rapidly with each iteration,
which can lead to numerical problems.  These are avoided similarly
to the procedure used in global strong-coupling flows in
position-space renormalization-group theory.  The physics described
by the tensor network is unaffected by an overall constant
multiplying each element in the tensor.  Thus at each step we factor
out the absolute value of a nonzero element, for example $T =
|T_{111}| \tilde{T}$, yielding a reduced tensor $\tilde{T}$. The
$T_{111}$ element is a convenient choice, since it remains nonzero
throughout the renormalization-group flow. The renormalization-group
transformation is then applied to the reduced tensor $\tilde{T}$,
giving a renormalized tensor $T^\prime$. Let us denote $T$ as
$T^{(0)}$ and $T^\prime$ as $T^{(1)}$.  The factorization and
renormalization-group transformation are iterated, so at the $n$th
step we have a tensor $T^{(n)} = |T^{(n)}_{111}| \tilde{T}^{(n)}$.
As we shall see below, with this added factorization step the
elements of $T^{(n)}$ tend to finite limiting values as $n \to
\infty$.

Additionally, keeping track of the factors $|T^{(n)}_{111}|$ during
the flow, as in standard position-space renormalization-group
theory, allows us to easily determine the free energy of the system
in the thermodynamic limit.\cite{Nauenberg}  Defining $G^{(n)}
\equiv \ln |T^{(n)}_{111}|$, the partition function is a contraction
of $2N$ tensors $T^{(0)}$, which we write schematically as $Z =
(T^{(0)})^{2N} = e^{2NG^{(0)}} (\tilde{T}^{(0)})^{2N}$. After a
single renormalization-group step, this becomes $Z = e^{2NG^{(0)}}
(T^{(1)})^{2N/3} = e^{2N (G^{(0)} + G^{(1)}/3)}
(\tilde{T}^{(1)})^{2N/3}$.  Thus after $n$ steps we find
\begin{equation}\label{eq:8}
Z = e^{2N\sum_{i=0}^n 3^{-i} G^{(i)}}
\left(\tilde{T}^{(n)}\right)^{2N/3^n}\,.
\end{equation}
The free energy per original site $\beta f = -N^{-1} \ln Z$ is then
\begin{equation}\label{eq:9}
\beta f = -2\sum_{i=0}^n 3^{-i} G^{(i)} -\frac{1}{N} \ln
\left(\tilde{T}^{(n)}\right)^{2N/3^n}\,.
\end{equation}
As $n \to \infty$ the elements of $\tilde{T}^{(n)}$ go to finite
fixed values, and thus for large $N$ the second term on the right in
Eq. \eqref{eq:9} becomes negligible. So the final result for the
free energy is $\beta f = -2\sum_{i=0}^\infty 3^{-i} G^{(i)}$.  In
practice the transformation is iterated for $n$ steps until the
value of $G^{(n)}$ has converged to the limit $G^{(\infty)}$ within
the desired numerical precision.  Then $\beta f \approx
-2\sum_{i=0}^{n-1} 3^{-i} G^{(i)} - 2 G^{(n)} \sum_{i=n}^\infty
3^{-i} = -2\sum_{i=0}^{n-1} 3^{-i} G^{(i)} -  3^{1-n} G^{(n)}$.

\begin{figure}
\includegraphics[scale=1]{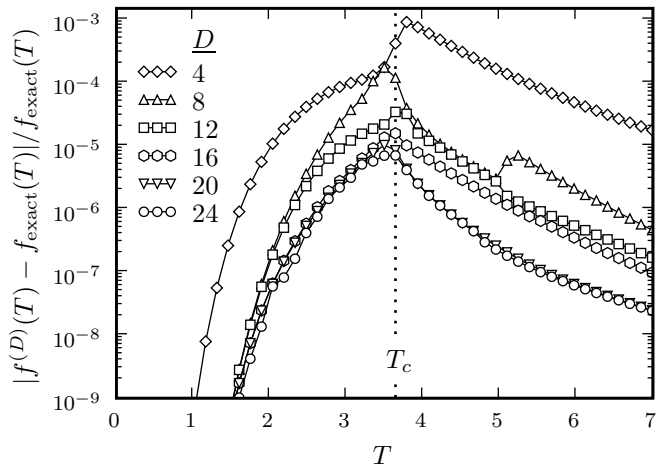}
\caption{The relative error of $f^{(D)}(T)$, the free energy per site
of the triangular lattice Ising model calculated using TRG at
various cutoffs $D=4$ to 24, compared to the exact free energy
$f_\text{exact}(T)$, as a function of temperature $T$.  The dotted
line shows the exact critical temperature $T_c = 4/\ln
3$.}\label{fig2}
\end{figure}

Even for small cutoffs $D$, the results of this free energy
calculation can be remarkably accurate, as seen in Fig. \ref{fig2},
which shows the relative error of $f^{(D)}(T)$, the free energy per
site using cutoff $D$, compared to the exact value
$f_\text{exact}(T)$.  The latter is calculated by numerical evaluation
of the integral solution in Ref. \cite{Wannier}.  Already at $D=4$ the
TRG free energy is within 0.09\% of the exact value at all
temperatures, and is considerably more accurate than this away from
the critical region near $T_c = 4/\ln 3$.  As noted in
Ref. \cite{LevinNave}, the TRG method is expected to behave
worst at criticality, and this is indeed what we see for all $D$, with
the relative error curves peaked near $T_c$.  Yet even here there is
significant improvement as we go to larger cutoffs.  At $D=24$, the
largest cutoff examined, the maximum error is 0.0007\%.  Overall,
going from $D=4$ to $D=24$ we get an improvement between two and three
orders of magnitude in the precision of the free energy result.

\begin{figure}
\includegraphics[scale=1]{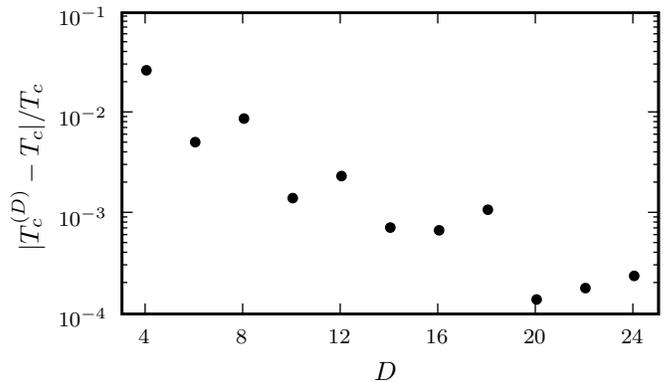}
\caption{The relative error of the critical temperature $T_c^{(D)}$
for the triangular lattice Ising model calculated from TRG flows
at various cutoffs $D = 4$ to 24, compared to the exact $T_c= 4/\ln
3$.}\label{fig3}
\end{figure}

\begin{figure}
\includegraphics[scale=1]{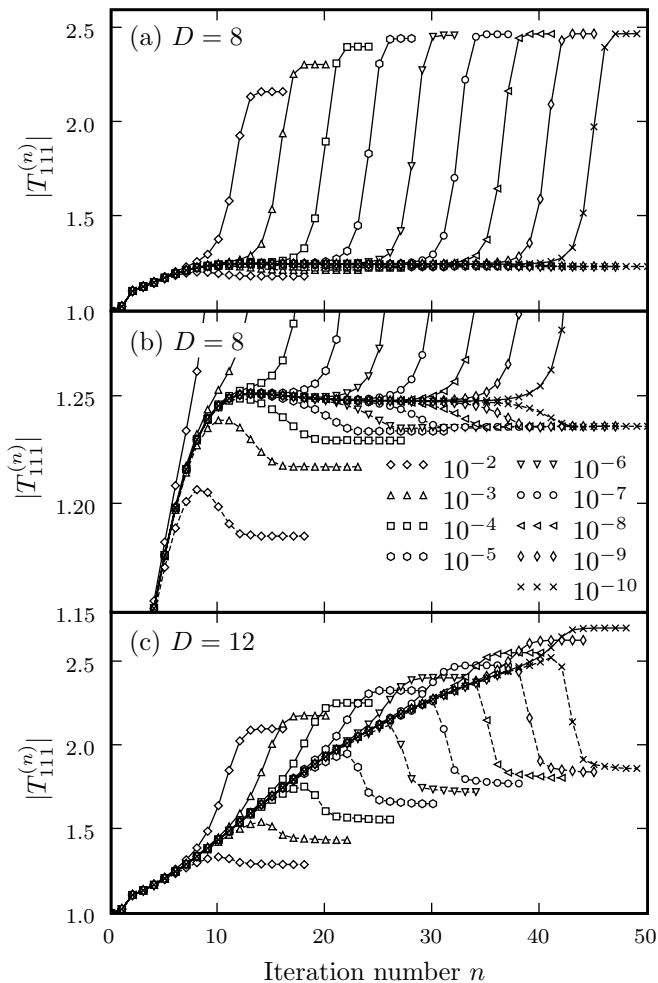}
\caption{The behavior of tensor element $|T^{(n)}_{111}|$ as a
function of iteration number $n$.  The different data sets correspond
to flows at different temperatures $T$ near $T_c^{(D)}$, with each
symbol in the legend corresponding to a value of $|\Delta T| =
|T-T_c^{(D)}|$ between $10^{-2}$ and $10^{-10}$.  For each $|\Delta
T|$ there are two data sets, one for $\Delta T >0$ connected by solid
lines, and the other for $\Delta T <0$ connected by dashed lines. For
clarity, the flows are only shown up to the iteration where they have
approximately converged to a fixed value.  (a) $|T^{(n)}_{111}|$ flows
with cutoff $D=8$. (b) Same as in (a), but zoomed in to see more
clearly the flows for $\Delta T < 0$.  (c) $|T^{(n)}_{111}|$ flows
with cutoff $D=12$.}\label{fig4}
\end{figure}

Further thermodynamic information can be gleaned by looking in
detail at the behavior of the renormalization-group flows.  The
elements $|T^{(n)}_{ijk}|$ go to finite fixed values
$|T^{\ast}_{ijk}|$ as $n \to \infty$, but unlike typical
position-space renormalization-group transformations, there are no
unique fixed points which act as a basins of attraction for the low-
and high-temperature phases.  Instead, at each different temperature
$T$ the system flows to a different fixed point
$|T^{\ast}_{ijk}(T)|$. However there is a way to distinguish the
low- and high-temperature phases, since each flows to a different
continuous surface of fixed points in the space of tensor elements.
There is a boundary temperature $T_c^{(D)}$, depending on the cutoff
$D$, such that for $T> T_c^{(D)}$ the flows tend to one fixed
surface, while for $T < T_c^{(D)}$ they tend to the other.  As $D$
becomes larger, $T_c^{(D)}$ gives a rapidly converging estimate of
the exact critical temperature $T_c = 4/\ln 3$.  Fig. \ref{fig3}
shows the relative error of $T_c^{(D)}$ compared to $T_c$ for
various $D$. While the decrease in error is not monotonic with $D$,
there is an overall trend which appears to be roughly exponential in
the cutoff, so that the error at $D=24$, $0.02\%$, is two orders of
magnitude smaller than at $D=4$.

\begin{figure}
\includegraphics[scale=1]{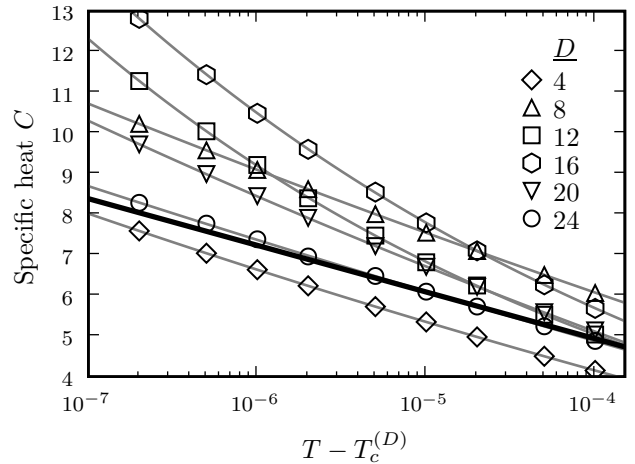}
\caption{Specific heat per site $C$ of the triangular lattice Ising
model, calculated using TRG for various cutoffs $D = 4$ to $24$
as a function of $T-T_c^{(D)}$ in the high-temperature phase near
criticality.  The gray curves superimposed on the data points are
best-fit curves of the form $A + B (T-T_c^{(D)})^{-\alpha}$, with
parameters $A$, $B$, and $\alpha$.  The thick black line is the exact
specific heat as a function of $T-T_c$.}\label{fig5}
\end{figure}

This segregation of the flows between two distinct fixed surfaces
can be seen directly in Fig. \ref{fig4}, which plots one tensor
element $|T_{111}^{(n)}|$ as a function of iteration number $n$ for
$D=8$ and $D=12$.  The flows are shown for different values of
$|\Delta T| = |T-T_c^{(D)}|$, from $10^{-2}$ to $10^{-10}$.  There
are two curves for each $|\Delta T|$, one corresponding to $T >
T_c^{(D)}$, and the other to $T < T_c^{(D)}$.  The two curves stay
close to one another for a number of iterations, but then veer off
in opposite directions and reach different fixed values
$|T^{(*)}_{111}|$. These fixed values change continuously as
$|\Delta T|$ is varied, as they map out a slice of the two fixed
surfaces corresponding to the low- and high-temperature phases.

As $|\Delta T|$ gets smaller, there is an interesting difference in
the flow behaviors of the $D=8$ and $D=12$ cases.  For $D=8$, as
seen in Fig. \ref{fig4}(a) and in more detail in Fig. \ref{fig4}(b),
the $\Delta T >0$ and $\Delta T <0$ curves are nearly horizontal
before turning away to their respective fixed surfaces.  This is
true in fact for all the $|T_{\alpha\beta\gamma}|$ elements, since
the flows are attracted to a unique critical fixed point, which we
denote $|T^{c*}_{\alpha\beta\gamma}|$.  The smaller the value of
$|\Delta T|$, the larger the number of iterations which are spent in
the vicinity of $|T^{c*}_{\alpha\beta\gamma}|$ before flowing to one
of the fixed surfaces.  Once we are in the vicinity of the critical
fixed point, we can isolate it to high precision using a
Newton-Raphson procedure.  The analysis of the fixed point proceeds
just as in a standard renormalization-group approach: we calculate a
recursion matrix, whose eigenvalues can be related to the critical
exponents of the system. To do this, let us denote the nonzero
elements of $|T^{c*}_{\alpha\beta\gamma}|$, not related by cyclical
symmetry, as $K_1$ through $K_m$.  In the case of $D=8$, $m = 80$.
For small perturbations away from the critical fixed point, the
number and locations of these nonzero elements stay the same after a
renormalization-group transformation, which allows us to numerically
evaluate the $m \times m$ recursion matrix $R_{ij} \equiv \partial
K^\prime_{i} /
\partial K_j$.  Writing the eigenvalues of $R$ in the form
$b^{y_i}$, $i = 1,\ldots,m$, we find only one eigenvalue where $y_i
> 0$, as expected at a critical fixed point.  This relevant
eigenvalue, which we denote $y_T$, is related to the specific heat
critical exponent $\alpha$ by $\alpha = (2y_T-2)/y_T$.  For $D=8$,
$y_T = 1.01543$, giving $\alpha = 0.03039$, which compare well to
the exact values of $y_T = 1$ and $\alpha = 0$. We can check this
result using an alternative approach, by calculating the specific
heat from derivatives of the calculated free energy for small
$T-T_c^{(D)}$, and we find that the singularity in the specific heat
agrees with the $\alpha$ derived from the thermal eigenvalue $y_T$
of the recursion matrix.

\begin{table}
\begin{tabular}{cccc}
\hline
\parbox{0.4in}{$D$} & \parbox{0.8in}{$T_c^{(D)}$} & \parbox{0.8in}{$T_c^{(D)}-T_c$} & \parbox{1in}{$y_T$} \\
\hline
4 & 3.73840 & $9.7\times 10^{-2}$ & 1.01543\\
8 & 3.60873 & $-3.2 \times 10^{-2}$ & 1.01543\\
12 & 3.64958 & $8.6 \times 10^{-3}$ & $1.0644 \pm 0.0048$\\
16 & 3.63847 & $-2.5 \times 10^{-3}$ & $1.0574 \pm 0.0022$\\
20 & 3.64147 & $5.1 \times 10^{-4}$ & $1.0165 \pm 0.0032$\\
24 & 3.64183 & $8.7 \times 10^{-4}$ & $1.0089 \pm 0.0075$\\
\hline
exact & 3.64096 & --- & 1\\
\hline
\end{tabular}
\caption{Critical properties calculated using the TRG approach,
at various cutoffs $D$, compared to the exact values in the last
row. The thermal eigenvalues $y_T$ for $D=4$ and 8 are calculated from
the recursion matrix evaluated at the critical fixed point.  For $D>8$
the estimate for $y_T$ is from the best-fit result to the specific
heat near $T_c^{(D)}$, as plotted in Fig. \ref{fig5}.}\label{tab1}
\end{table}

For the six values of the cutoff $D$ where we investigated the
near-critical flows in detail, $D = 4, 8, 12, 16, 20, 24$, we were
able to isolate the critical fixed point for $D=4$ and $8$, namely
in 24- and 196-dimensional flow spaces. Despite the different
dimensionalities of the flow spaces, both cases yielded the same
eigenvalue $y_T$, within the precision of 5 decimal places.  Higher
values of $D$ showed very different flow behaviors, as exemplified
in Fig. \ref{fig4}(c) for $D=12$. Here the $\Delta T >0$ and $\Delta
T < 0$ curves for $|T_{111}^{(n)}|$ do not stay nearly horizontal
before diverging to the fixed surfaces: regardless of how small
$|\Delta T|$ is made, the flows do not gravitate toward a unique
critical fixed point, but map out a continuous spectrum of points
which attract the flows before the two curves spread out.  Using an
arbitrary precision version of the TRG algorithm implemented in {\it
Mathematica}, we checked $|\Delta T|$ values as small as $10^{-43}$
without finding convergence toward a critical fixed point.
Nevertheless, in these cases for $D > 8$ we can still extract the
critical behavior, by resorting to the alternative approach
mentioned earlier: looking directly at the specific heat $C$ per
site near $T_c^{(D)}$ as obtained from the numerical derivative of
the calculated free energy. We plot $C$ within the high-temperature
phase for various $D$ in Fig. \ref{fig5}, for $T-T_c^{(D)}$ between
$10^{-7}$ and $10^{-4}$. For each $D$ the data points are fit to a
function $A+B(T-T_c^{(D)})^{-\alpha}$, which provides an accurate
description of the singularity, and the best-fit value of $\alpha$
is used to determine the thermal eigenvalue $y_T$.  The results are
listed in Table \ref{tab1}.  For $D=12$, $y_T = 1.0644 \pm 0.0048$,
worse than our fixed-point determined values at $D=4$ or $8$, but
this value improves as $D$ is increased, reaching $y_T = 1.0089 \pm
0.0075$ at $D=24$. Moreover, as seen in Fig. \ref{fig5}, our
calculated $C$ curve for $D=24$ nearly overlaps the exact $C$ as a
function of $T-T_c$, which diverges with a logarithmic singularity
at the critical point.

\section{Conclusion}

In summary, we have seen that the flows of the tensor elements in
the TRG transformation can be used to extract the phase diagram
structure and critical behavior of a classical two-dimensional
lattice Hamiltonian.  For the triangular lattice Ising model, the
low- and high-temperature phase regions are basins of attraction for
two distinct surfaces of fixed points.  The boundary between these
basins defines a critical temperature $T_c^{(D)}$, dependent on the
TRG cutoff $D$.  At small cutoffs such as $D = 4$ and 8, the flows
near the boundary between the basins are controlled by a critical
fixed point, while at higher $D$ the flows show more complicated
behavior, never converging at a unique point.  In the former case
the thermal exponent $y_T$ is found from the eigenvalues of the
recursion matrix at the critical fixed point, while in the latter we
can deduce the exponent from the scaling of the calculated specific
heat near $T_c^{(D)}$.  The free energy at all temperatures
systematically converges to the exact Ising result with increasing
$D$, particularly fast away from the critical region.  For the
critical properties the improvement is not monotonic in $D$, but
both the critical temperature $T_c^{(D)}$ and the exponent $y_T$
tend toward the exact Ising values at larger cutoffs.  With very
modest computational effort, the TRG method provides an accurate
portrait of global phase diagram characteristics. It thus warrants
further study, both on applications to other two-dimensional lattice
models, and possible generalization to systems with quenched
randomness \cite{Falicov} and/or higher spatial dimensions.

\begin{acknowledgments}
This research was supported by the Scientific and Technical Research
Council (T\"UB\.ITAK) and by the Academy of Sciences of Turkey.
\end{acknowledgments}


\begin{thebibliography}{}
\bibitem{LevinNave} M. Levin and C.P. Nave, arXiv:cond-mat/0611687.
\bibitem{NvL} Th. Niemeijer and J.M.J. van Leeuwen, Phys. Rev. Lett. {\bf 31}, 1411 (1973).
\bibitem{Migdal} A.A. Migdal, Zh. Eksp. Teor. Fiz. {\bf 69}, 1457 (1975) [Sov.
Phys. JETP {\bf 42}, 743 (1976)].
\bibitem{Kadanoff} L.P. Kadanoff, Ann. Phys. (N.Y.) {\bf 100}, 359 (1976).
\bibitem{Kadanoff1} L.P. Kadanoff, Phys. Rev. Lett. {\bf 34}, 1005 (1975).
\bibitem{Kadanoff2} L.P. Kadanoff, A. Houghton, and M.C. Yalab\i k,
J. Stat. Phys. {\bf 14}, 171 (1976).
\bibitem{Berker} A.N. Berker and M. Wortis, Phys. Rev. B {\bf 14}, 4946 (1976).
\bibitem{MarkovShi} I. Markov and Y. Shi, arXiv:quant-ph/0511069.
\bibitem{ShiDuanVidal} Y. Shi, L. Duan, and G. Vidal, Phys. Rev. A {\bf 74}, 022320 (2006).
\bibitem{Takagi} T. Takagi, Japan. J. Math. {\bf 1}, 83 (1925).
\bibitem{LAPACK} E. Anderson {\it et. al.}, {\it {LAPACK} Users' Guide} (SIAM, Philadelphia, PA, 1999), 3rd ed.
\bibitem{Nauenberg} M. Nauenberg, J. Math. Phys. {\bf 16}, 703 (1975).
\bibitem{Wannier} G.H. Wannier, Phys. Rev. {\bf 79}, 357 (1950); {\it erratum}: Phys. Rev. B {\bf 7}, 5017 (1973).
\bibitem{Falicov} A. Falicov, A.N. Berker, and S.R. McKay, Phys. Rev. B {\bf 51}, 8266
(1995).

\end{thebibliography}
\end{document}